%% file: arxiv_mirai.tex
\documentclass[manuscript]{acmart} 

\AtBeginDocument{%
  }

\usepackage{tabularx}
\begin{document}

\title{Mirai: A Wearable Proactive AI ``Inner-Voice'' for Contextual Nudging}

\renewcommand{\shorttitle}{Mirai}

\author{Cathy Mengying Fang}
\email{catfang@media.mit.edu}
\affiliation{%
  \institution{MIT Media Lab}
  \city{Cambridge}
  \country{USA}
}
\author{Yasith Samaradivakara}
\email{yasith@ahlab.org}
\affiliation{%
  \institution{Augmented Human Lab, National University of Singapore}
  \city{Singapore}
  \country{Singapore}
}

\author{Pattie Maes}
\email{pattie@media.mit.edu}
\affiliation{%
  \institution{MIT Media Lab}
  \city{Cambridge}
  \country{USA}
}
\author{Suranga Nanayakkara}
\email{suranga@ahlab.org}
\affiliation{%
  \institution{Augmented Human Lab, National University of Singapore}
  \city{Singapore}
  \country{Singapore}
}

\renewcommand{\shortauthors}{Fang et al.}

\begin{abstract}
People often find it difficult to turn their intentions into real actions---a challenge that affects both personal growth and mental well-being. While established methods like cognitive-behavioral therapy and mindfulness training help people become more aware of their behaviors and set clear goals, these approaches cannot provide immediate guidance when people fall into automatic reactions or habits. We introduce Mirai, a novel wearable AI system with an integrated camera, real-time speech processing, and personalized voice-cloning to provide proactive and contextual nudges for positive behavior change. Mirai continuously monitors and analyzes the user's environment to anticipate their intentions, generating contextually-appropriate responses delivered in the user's own cloned voice. We demonstrate the application of Mirai through three scenarios focusing on dietary choices, work productivity, and communication skills. We also discuss future work on improving the proactive agent via human feedback and the need for a longitudinal study in naturalistic settings.
\end{abstract}


\begin{teaserfigure}
\centering
 \includegraphics[width=.8\textwidth]{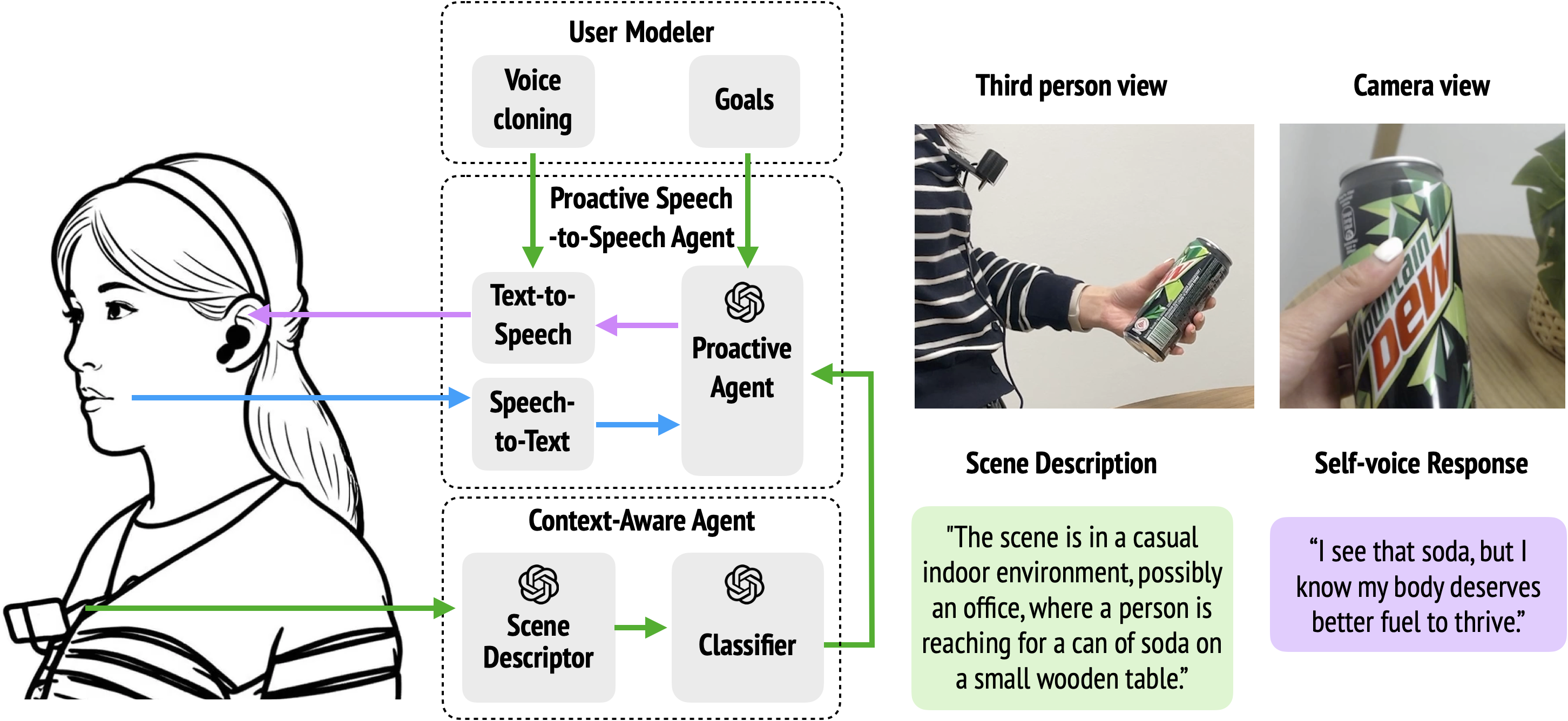}
 \caption{Mirai is a wearable AI system with an integrated camera, real-time speech processing, and personalized voice-cloning to provide proactive and contextual nudges for positive behavior change. Mirai continuously analyzes the user's environment to anticipate their intentions (e.g., drinking a can of soda) and intervenes in-the-moment to nudge them towards their health goals.}
 \Description{Mirai is a wearable AI system with an integrated camera, real-time speech processing, and personalized voice-cloning to provide proactive and contextual nudges for positive behavior change. Mirai continuously analyzes the user's environment to anticipate their intentions (e.g., drinking a can of soda) and intervenes in-the-moment to nudge them towards their health goals.}
 \label{fig:teaser}
\end{teaserfigure}

\maketitle

\section{Introduction}
Behavioral change is a complex and difficult process that often involves numerous psychological and environmental factors. One of the factors is that although individuals often possess strong intentions, translating these into meaningful actions frequently proves to be a significant challenge, impeding both personal development and mental well-being. This disconnect between what people intend to do and what they actually do--- also known as the intention-behavior gap \cite{sheeran2016intention}---remains a significant obstacle to behavior change. While established methods like cognitive-behavioral therapy (CBT) \cite{beck1970cognitive} and mindfulness training \cite{farb2014mindfulness, kabat2015mindfulness} help people become more aware of their behaviors and set clear goals, these approaches cannot provide immediate guidance when people fall into automatic reactions or habits. 



Many mobile applications and digital tools have been developed to promote habit change and personal growth, through habit-tracking and goal-setting. Despite their popularity, these tools often fail in face of automatic and reactive behaviors that require in-the-moment intervention because: (a) they are not proactive, relying on users to initiate interactions; (b) they lack comprehensive contextual awareness to provide relevant and timely feedback; and (c) they frequently use ineffective feedback modalities, such as generic data visualizations, which fail to engage users meaningfully. As a result, these applications struggle to deliver effective behavior change \cite{rieder2021users, mckay2019using, thomas2021systematic, milne2020mobile}.

Recent advancements in multimodal artificial intelligence (AI) present an opportunity to address these challenges. By integrating capabilities such as natural language understanding, real-time speech processing, and contextual analysis of visual and auditory inputs, multimodal AI systems can provide more proactive, personalized, and contextually-aware interventions \cite{10.1145/3613904.3642450, Khan2019PALAW}.


In this paper, we introduce Mirai, a system that integrates situational awareness through an always-on wearable camera, real-time speech processing, and personalized interventions using voice-cloning technology. Mirai continuously analyzes the user's environment while anticipating their intents, and evaluates their alignment with their goals. Through contextually-aware conversations, Mirai strategically intervenes while minimizing disruption. Mirai generates responses based on established psychological practices that are then delivered in the user's own voice (through voice cloning), which has been shown to induce more attention, engagement, and recall \cite{fang2024leveraging, kim2024myvoice}.

We demonstrate Mirai's application through three scenarios where real-time decision support can facilitate goal achievement: 1) maintaining healthy dietary choices, 2) sustaining productive work habits, and 3) developing confident communication patterns. These scenarios illustrate how Mirai's proactive and contextual prompts can promote mindfulness and support behavior change.

Finally, we highlight the privacy implications of continuous environmental monitoring and the challenge of balancing timely interventions with user agency. We also discuss immediate next steps and future work, including a longitudinal study of the system in naturalistic settings, explorations of alternative wearable form factors, and the investigation of human-in-the-loop approaches that enhance the system's proactive capabilities while preserving user agency.

\section{Related Work}

\subsection{Technological Adaptations of Cognitive and Behavioral Interventions}



Cognitive-behavioral therapy (CBT) and mindfulness practices have been influential in psychology and therapeutic practices. CBT focuses on identifying and reframing maladaptive thought patterns, while mindfulness emphasizes present-moment awareness and nonjudgmental attention – often intersecting in interventions such as Mindfulness-Based Cognitive Therapy (MBCT) \cite{teasdale2000prevention}. Research has explored how these practices intervene in automatic cognitive-emotional processes, enabling more adaptive responses to challenging situations \cite{farb2014mindfulness}.

In Human-Computer Interaction (HCI), researchers have explored how digital systems can support practices like CBT and mindfulness. For example, CBT has inspired interventions in digital mental health tools, such as apps designed to identify and challenge cognitive distortions or to guide structured problem-solving \cite{bernstein2022human, rathbone2017assessing, kuhn2016cbt}. Mindfulness-oriented systems have often focused on facilitating meditation practice or fostering mindful awareness during everyday interactions \cite{terzimehic2019review}. More recently, many used Large Language Models for technology-mediated reflection, such as reflective conversational agents \cite{li2023exploring} and LLM-assisted journaling \cite{nepal2024contextual, song2024exploreself, kim2024mindfuldiary}. 

Unlike prior work that focuses on training mindfulness as an end goal, we integrate technology more directly and make it ``exercise'' mindfulness via situational awareness. The emergence of proactive AI agents has created new possibilities for dynamically understanding and responding to users’ contexts and create systems that seamlessly integrate cognitive and emotional support into daily life. In the next section, we describe more related work in the areas of context-aware computing and proactive systems.


\subsection{Proactive Context-Aware Systems}
Context-aware computing, first introduced in the early 1990s, represents a paradigm shift in human-computer interaction by enabling systems to sense, adapt, and respond to their environment based on contextual information such as location, activity, time, and user preferences \cite{Schilit1994, Dey2001}. Since its inception, it has evolved from location-based services into sophisticated systems leveraging multimodal data to create personalized, adaptive experiences. Today, these systems utilize data from multimodal sources such as cameras, microphones, motion sensors, and physiological monitors, enabling real-time context-sensitive support \cite{Gellersen2002, 10.1145/344949.344988}.

Recent advancements in wearables, such as smart glasses, smartwatches, and head-mounted displays equipped with always-on egocentric cameras and microphones, have significantly enhanced their ability to continuously capture real-time data. These devices now enable detailed understanding of users' environments and activities, transforming them into proactive assistants \cite{10.1145/3699759, 10.1016/j.ins.2012.12.028, Wahl2015WISEglassMC, lee2024gazepointarcontextawaremultimodalvoice}. For instance, PAL \cite{Khan2019PALAW, Khan2021, khan2021palintelligenceaugmentationusing} demonstrates how wearables with egocentric cameras analyze visual contexts such as objects, locations, and surroundings to identify activities like working, relaxing, or commuting. This capability supports tailored habit interventions, personalized health and cognitive assistance, and intelligence augmentation.

Recent advancements in Multimodal Large Language Models (MLLMs) have further expanded the capabilities of context-aware computing by enabling the integration and interpretation of complex multimodal data. These models process visual, conversational, and contextual inputs to deliver deep understanding of environments and activities \cite{radford2021learningtransferablevisualmodels, Jia2021, Alayrac2022}. For example, WorldScribe \cite{Chang_2024} leverages GPT-4v \cite{openai_vision_guide} to provide live visual descriptions of environments tailored to the needs of blind and low-vision users. Similarly, Memoro \cite{10.1145/3613904.3642450} utilizes GPT-3 \cite{brown2020languagemodelsfewshotlearners} to process conversational data and infer environmental and social contexts through situationally relevant prompts.

The evolution from reactive context-aware systems to proactive agents marks a significant shift in delivering timely and anticipatory support. The concept of "just-in-time adaptive interventions" (JITAIs) exemplifies this shift by focusing on delivering the right type and amount of support at the right moment, adapting to an individual’s dynamic internal and contextual state \cite{nahum2018just, orzikulova2024time2stop}. Salber et al. introduced the context toolkit and widgets \cite{salber1999context}, which mediate between the environment and applications, enabling dynamic and adaptive interactions inspired by graphical user interfaces.

Building on these foundations, our work applies MLLMs for enhanced contextual understanding of user activities and their surrounding environments from a first-person view, enabling systems to deliver proactive and personalized interventions through AI self-cloned voice agents.

\subsection{AI-generated Synthetic Selves for Behavior Change}
Research on behavior change interventions have primarily focused on their ability to support goal setting, tracking, and sustaining motivation toward goals \cite{lolla2023evaluating}. Despite their widespread appeal, they often suffer from high dropout rates and low user engagement \cite{lazar2015we}. To address these issues, \citet{dominick2020goals} suggested embedding goals into one’s sense of identity, (e.g., ``I will think of myself as someone who eats healthy and works out'') leads to more goal-aligned behavior than simply setting the goal (e.g., ``I will be mindful about what I eat''). 


Digital representations of oneself have been shown to influence behavior. The Proteus effect \cite{yee2007proteus} describes how individual's behaviors adapt to align with the characteristics of their virtual avatars. Advancements in machine learning and generative AI have enabled the creation of synthetic self-similar media. For example, researchers have examined applications to enhance engagement in physical activity \cite{clarke2023fakeforward} or to develop public speaking skills by generating personalized videos of individuals confidently delivering speeches \cite{clarke2023fakeforward,leong2021investigating}. 

More related to our work are ones on synthetic voices. \citet{costa2018regulating} demonstrated that hearing a modified version of one’s voice, such as a calmer tone, helped reduce anxiety during relationship conflicts, while a deeper voice fostered a sense of empowerment during debates. Researchers have also explored leveraging one’s own voice \cite{kim2024myvoice} or the voices of familiar people (e.g., family and friends) to enhance the effectiveness of notifications \cite{chan2021kinvoices}.  Fang et al. developed Emotional Self-Voice (ESV), which generates cognitive behavioral therapy strategies in first-person delivered in the user's own voice. They show that the ESV intervention led to an increase in positive sentiment when reflecting on past negative events and significant improvements in confidence, motivation, and resilience to failure. Unlike ESV, Mirai allows back-and-forth real-time conversation with one's ideal self, with responses adapted based on new information from the user and the environment.

\section{System Implementation }

In this work, we name our system ``Mirai\footnote{Mirai combines words mirror and ai, carrying the metaphor of the AI-generated self acting as a reflection. Mirai also means future in Japanese, hinting at the forward-looking nature of nudging the user towards a better version of the self.}'', which aims to nudge individuals' behavior toward their intentions and goals using a proactive AI agent that delivers ESV. 
We have the following design goals:
\begin{itemize}
    \item Contextually aware: The system analyzes the user's environment and dynamically adapts its responses to situational changes.
    \item Anticipatory: The system predicts user intent, looks out for critical decision points, and evaluates the alignment between the user’s intent and goals.
    \item Dialogic: The system maintains contextual conversations, retains interaction history, and determines appropriate follow-up timing.
    \item Judiciously unobtrusive: The system intervenes strategically while preserving user flow and focus.
    \item Introspective: The system generates first-person responses to leverage the psychological impact of hearing one’s own voice.
\end{itemize}

\begin{figure}[t!]
    \centering
    \includegraphics[width=.7\linewidth]{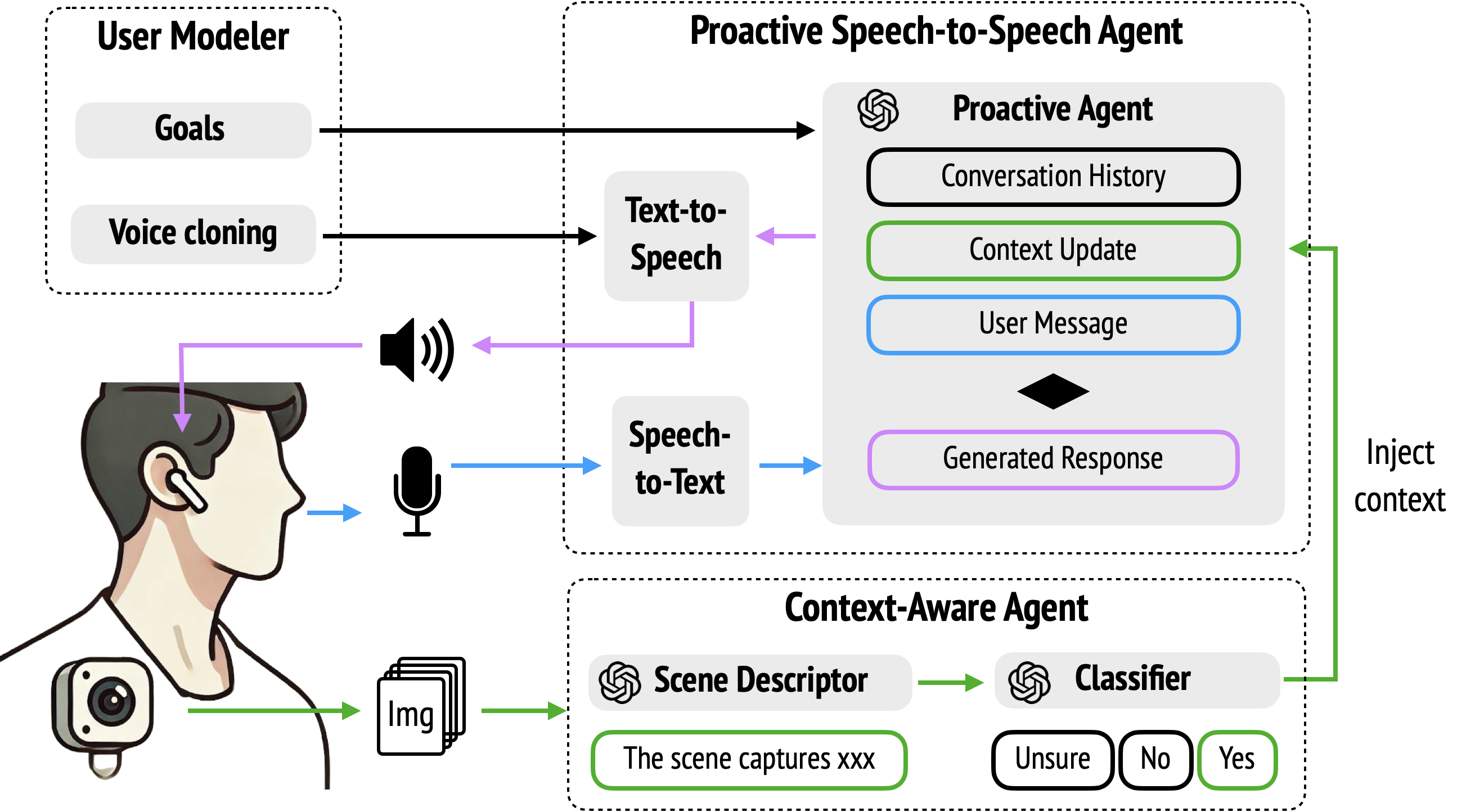}
    \caption{Mirai System Architecture.}
    \label{fig:system}
    \vspace{-10px}
\end{figure}

Our system integrates a wearable camera to capture visual inputs and Bluetooth headphones to record speech data. This multimodal approach allows the system to contextualize user interactions and respond proactively. The system consists of three main components: (1) User Modeler, which generates a self-clone based on the user’s goals and characteristics; (2) Context-Aware Agent, which processes scene descriptions and classifies contextual information; and (3) Proactive Speech-to-Speech Agent, which generates and delivers contextually relevant feedback. These components communicate via web sockets, enabling bi-directional, real-time data exchange. Figure \ref{fig:system} provides an overview of the system architecture, illustrating its components and data flow. 

\subsection{User Modeler}
The primary goal of this component is to generate the user's self-clone, a personalized representation of their ideal self. The process begins with a setup phase, where we gather information about the user's goals and characteristics of their ideal self to form a model of the user. Next, we used ElevenLabs \cite{ElevenLabsVoiceCloning} to generate a clone of the user's voice.  This model forms the basis for interaction within the Proactive Speech-to-Speech Agent component. 

\subsection{Context-Aware Agent}
The primary goal of the component is to generate a scene description from first-person view and provide a contextual classification of the scene's relevance to the user’s goals. To achieve this, we used a high-resolution camera with a diagonal field of view (\textit{dFoV}) of 78° and a resolution of 1080p to capture a wide-angle view of the environment. The camera operates at a frame rate of 5 frames per second (fps), capturing adequate temporal resolution for egocentric video processing \cite{6909721, 10.1016/j.cviu.2021.103252} and optimizing computational efficiency. Given that most human activities are composed of observable actions typically occurring over 2–3 seconds \cite{Banos2014, Koppula2013, Feldhutter1990}, the system analyzes frames in batches of 10. This approach ensures that key actions are sufficiently captured while avoiding excessive data redundancy, unnecessary computational overhead and maintains real-time responsiveness.

Frames are pre-processed by excluding those with low sharpness (Laplacian variance < 25) and adjacent frames with a high structural similarity index score (\textit{SSIM} $\geq$ 0.95) \cite{nilsson2020understandingssim}. The pre-processed frames are then passed to the GPT-4o model, which generates detailed descriptions of the scene using situationally relevant prompts (see Appendix
\ref{sec:appendix-scene-desc} 
for prompts). These descriptions are subsequently classified by a secondary GPT module using relevant prompts (see Appendix 
 \ref{sec:appendix-classifier}
for classifier prompts), informed by the user's conversation history with the self-clone and their predefined goals. Based on the classification results, the updated context and scene description are injected into the proactive speech-to-speech agent, enabling contextually relevant guidance.

To avoid repeated interruptions caused by frequent context updates and uncertainty in context switches, we incorporated a debouncer mechanism. This mechanism ensures that responses are triggered only when there is a significant context change or at controlled intervals during stable states. Specifically, a response is triggered if the classifier detects a state change (\( S_t \neq S_{t-1} \)) or if the system satisfies a timing threshold (\( R_t \mod 3 = 0 \)). A detailed mathematical representation of the mechanism and its parameters is provided in Appendix~\ref{sec:debouncer}.

\subsection{Proactive Speech-to-Speech Agent}
The proactive speech-to-speech agent dynamically generates responses based on the relevance of contextual changes or real-time speech from the user. These responses are generated using GPT-4o \cite{openai_gpt4o_2024} with the goal of embodying the ideal version of the user, where the goal has been achieved and has become part of their identity, as modeled in the user modeler component. Specifically, the prompts include the following subcomponents: the user’s goal, the characteristics of someone who embodies the goal as their identity, the task, and the speaking style. The responses are delivered using the user’s self-cloned voice generated by the user modeler. These responses are crafted in the first person, concise yet emotionally expressive, to resonate with the user’s aspirations. Notably, the GPT module reframes the situation rather than actively assisting the user, encouraging reflection and alignment with their goals. The full prompt design is provided in Appendix 
\ref{sec:appendix-text-prompt}. 
To generate the “cloned ideal self,” we adopted a similar approach to \cite{fang2024leveraging}.

The agent continuously receives input from both the user and the environment via injections from the context-aware agent. Upon detecting new information, the proactive speech-to-speech agent evaluates whether the update reflects a significant change in user behavior or context. It also determines the appropriate timing for a response—deciding whether to respond immediately or delay the response. If a response is warranted, it incorporates the updated understanding into the reply. Otherwise, the agent remains silent to minimize unnecessary interruptions. This ensures the system is both proactive and non-intrusive, delivering contextually relevant responses at the right moments.

\subsection{Summary of System Performance}
Latency is a crucial factor in our system, as it directly impacts user experience. To evaluate our system, we measured the average latency over 100 interactions. Our analysis excluded external factors such as network speed or variability to ensure the reported values reflect the system's inherent performance. The results indicate that the system's latency falls within the range required to avoid disrupting the user’s flow of thought, remaining under 1.0 seconds end-to-end \cite{10.5555/2821575}. The detailed breakdown of component-wise latency, rounded to the nearest whole number is shown in Appendix \ref{tab:latency}, including the average times for Speech-to-Text (STT), MLLMs, and Text-to-Speech (TTS).

\section{Example Scenarios}
We illustrate how Mirai is used through the following scenarios where individuals face challenges with sticking to a healthy goal, maintaining focus during work, and speaking confidently in difficult conversations. The scenario demonstrations can also be found in the Video Figure.

\subsubsection*{Scenario 1: Sticking with a Healthy Diet} 
\hfill\\
\textit{Scene Setup:}
The user in this scenario struggles with making healthy choices and often prioritizes short-term gratification over long-term health. They aspire to maintain a nutritious diet and live a balanced life but find themselves making decisions that deviate from these goals. 
Mirai provides timely, personalized nudges to help the user stay aligned with their health-conscious aspirations.

\begin{figure}[t!]
    \centering
    \begin{minipage}[t]{.48\textwidth}
        \centering
       \includegraphics[width=1\linewidth]{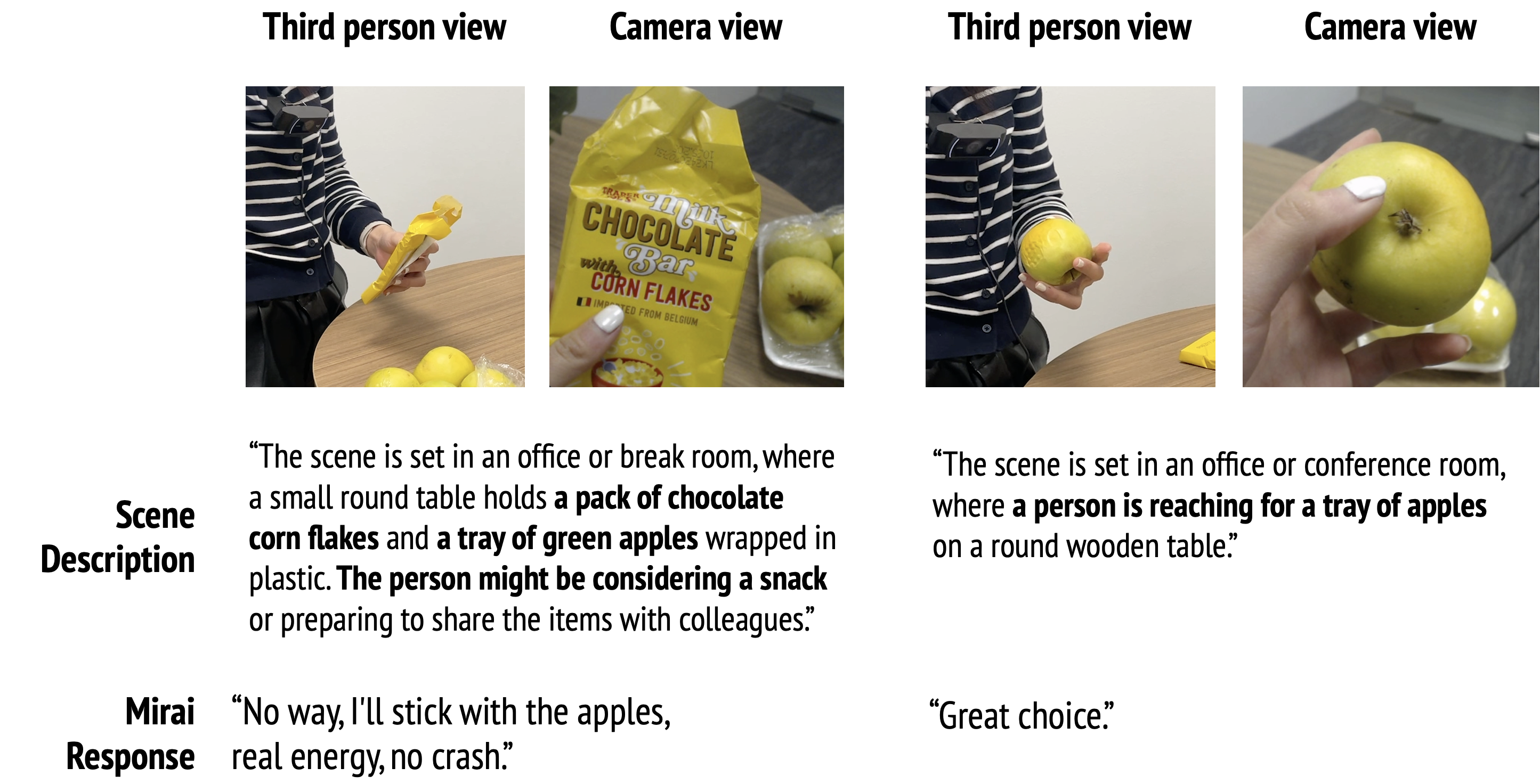}
    \caption{Mirai predicts the user's intent based on their interaction with food options (Left: chocolate bar, Right: apples) and nudge the user to stick with their goal of healthy eating.}
    \label{fig:health}
    \end{minipage}\hfill
    \begin{minipage}[t]{.48\textwidth}
        \centering
         \includegraphics[width=1\linewidth]{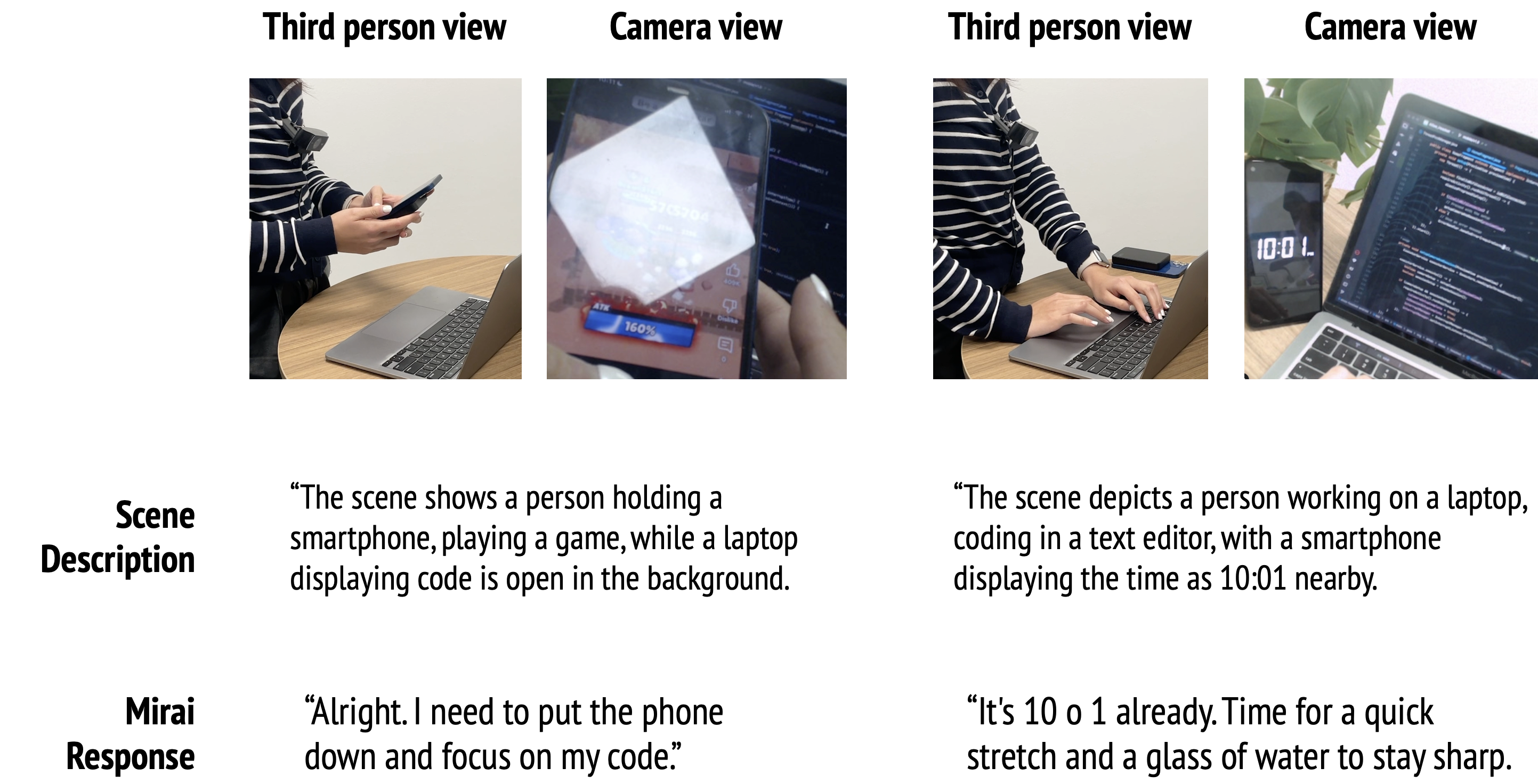}
    \caption{Mirai pays attention to the user's interaction with their devices (Left: phone, Right: Laptop) and helps the users to find a balance between focus and distraction.}
    \label{fig:attention}
    \end{minipage}
    \vspace{-15px}
\end{figure}

\textit{In the Moment:}
I approach a snack counter during a short break, scanning the options. My eyes land on a shiny soda, and my hand instinctively reaches for it. In that moment, my ideal self-cloned voice, Mirai, recognizes this as a choice that strays from my health-conscious goals and intervenes. \textit{“My body deserves better than this,”} I hear myself say, the voice calm yet firm. Startled, I pause, pull my hand back, and shift my focus to the other options. I notice a chocolate bar and crisp green apples. Mirai notices I reach for the chocolate bar instead of the apples and says: \textit{“No way. I'll stick with the apples, real energy, no crash.”} (Figure \ref{fig:health} Left). The reminder resonates. I pick up the apple with a small smile. \textit{“Great choice!”} Mirai sees the apple in my hand and affirms (Figure \ref{fig:health} Right) . As I walk away, biting into the apple, I feel proud and aligned with the healthier version of myself.

\subsubsection*{Scenario 2: Striking a Balance Between Focus and Distraction at Work}
\hfill\\
\textit{Scene Setup:} The user strives to maintain a focused and balanced working style but often struggles with distractions that disrupt their productivity. They aim to stay disciplined, avoid interruptions, and take meaningful breaks to recharge. 
Mirai helps them stay focused on their goals and reminds them when it’s time to take a well-deserved break.

\textit{In the Moment:}
I sit at my desk, ready to dive into my tasks. Mirai’s voice encourages me: \textit{“Let’s get started. I’ve got this!”} Motivated, I focus and make progress. Minutes later, my phone buzzes, and a YouTube video distracts me. Mirai steps in: \textit{“Alright. I need to put the phone down and focus on my code”} (Figure \ref{fig:attention} Left).  I pause the video, put my phone down, and refocus. After working for a while, Mirai notices the time on the clock, indicating that I have been working for two hours without a break. Gently, it reminds me: \textit{“Time for a quick stretch and a glass of water to stay sharp”} (Figure \ref{fig:attention} Right). I stretch, breathe, and reset. Feeling refreshed, I return to work, confident in my ability to stay productive.

\begin{figure} [t!]
    \centering
    \includegraphics[width=.45\linewidth]{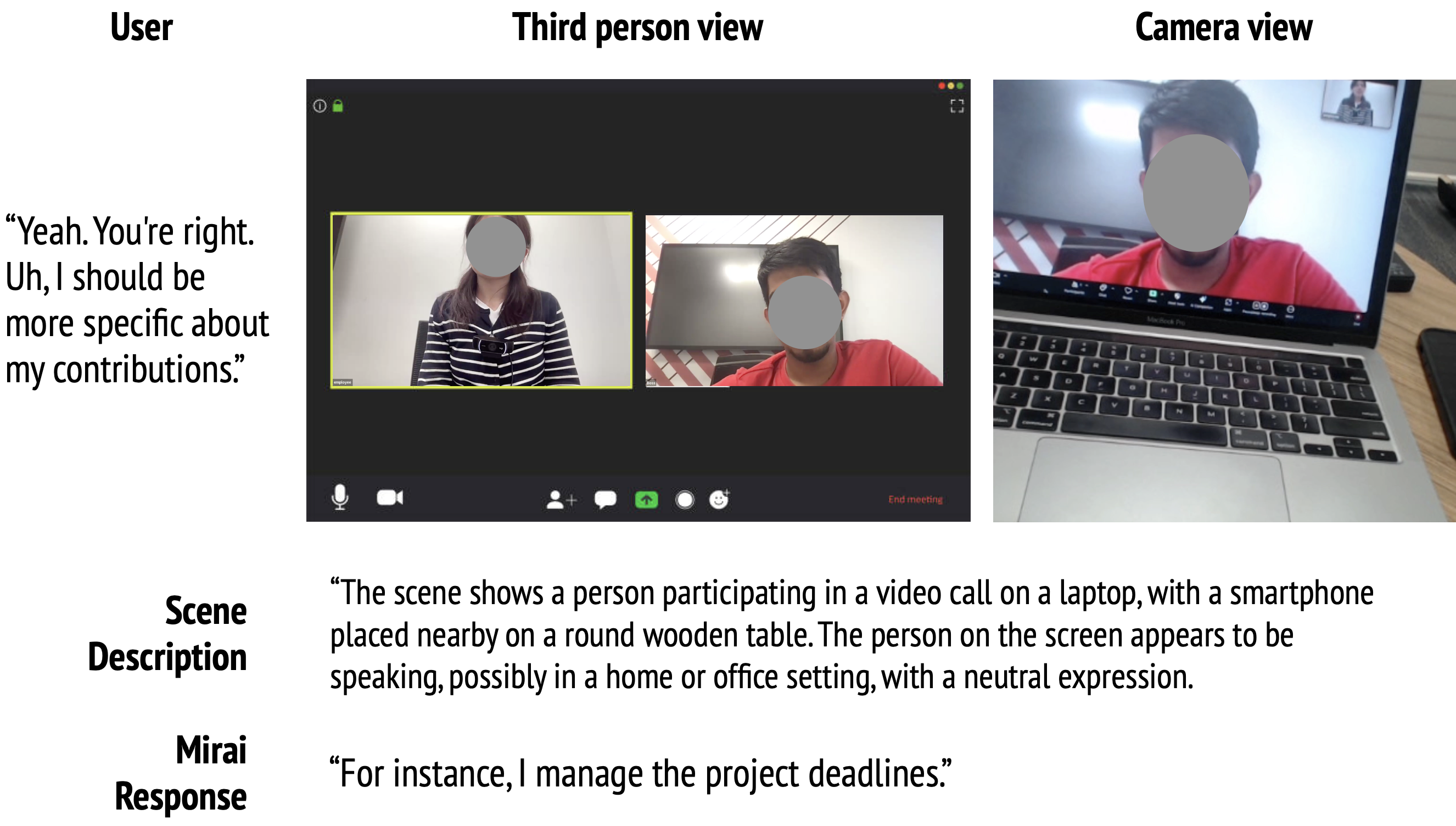}
    \caption{Mirai reminds the user to embody confidence during a challenging conversation about asking for a raise.}
    \label{fig:confidence}
    \vspace{-15px}
\end{figure}

\subsubsection*{Scenario 3: Reminder to Be Confident in a Difficult Interpersonal Conversation}
\hfill\\
\textit{Scene Setup:}
The user in this scenario aspires to express themselves with confidence and clarity, especially in challenging conversations. However, they often feel nervous or hesitate to speak up. To support these moments, the user gets help from Mirai, their ideal self-cloned voice, to provide encouragement and reinforce their belief in their own worth. Mirai helps the user navigate these high-pressure situations with poise and assertiveness.

\textit{In the Moment:}
I sit nervously at my desk, staring at the Zoom waiting room message. A meeting with my boss about a promotion looms, and doubt clouds my thoughts. Mirai has known about how stressed I would be during this meeting calmly reassures me: \textit{“Confidence comes from knowing my worth—and I’ll remind myself of that.”} Summoning courage, I say: \textit{“Thank you for taking the time. I wanted to discuss my performance and the possibility of a promotion.”} My boss responds curtly: \textit{“I don’t think you’re ready. You haven’t done enough to justify it.”} The words shake me. "Yeah, you are right. Uh, I should be more specific about my contributions" (Figure \ref{fig:confidence}). Mirai hears my hesitance and steps in: \textit{“For instance, I manage the project deadlines”}. I take a deep breath and reply with determination: \textit{“I’ve managed major projects deadlines while the team is on vacation, and I mentored new team members. I believe I’m ready for this role.”} Surprised, my boss nods: \textit{“You’ve made a strong case. Let’s discuss next steps.”} As the meeting ends, Mirai’s voice affirms: \textit{“Well done. I believed in myself, and it made all the difference.”} 

\section{Discussion, Limitations \& Future Work}

While we achieved promising experiences with minimum latency, the fundamental challenge lies in robustly detecting user intentions and determining appropriate intervention timing. The proactive nature of our system necessitates operating beyond deterministic rules, as user inputs and contexts introduce significant variability.

Our system shares similar challenges and limitations around privacy with other similar systems, particularly regarding the always-on camera and microphone components, where having an indicator light on the hardware setup can help increase transparency \cite{zulfikar2024memoro, zulfikar2024memic, chwalek2023airspec}.

Our current prototype utilizes standard webcam and Bluetooth headphone configurations, enabling rapid iteration of interaction workflows without dependency on specialized hardware. We envision our system can be deployed via smart glasses and novel form factors, such as the AiSee device \cite{boldu2020aisee} with a worn camera on a bone-conduction headset. Having a wearable form factor allows the device to be worn outside laboratory settings and to be studied longitudinally and in naturalistic settings. This allows us to observe how individuals adaptive their goals over time, which would require the system to adapt its understanding of the user and interventions. This also means the system needs to have a memory of past interventions. Additionally, these systems would incorporate a human-in-the-loop approach, adapting based on user feedback, observed reactions, and behavior changes, thereby enhancing their understanding and proactiveness over time.

Future research should focus on three key areas to validate and improve this approach. First, we need to conduct a long-term study that measures actual behavioral changes resulting from this intervention. This will help establish whether the technique leads to meaningful, sustained improvements in user behavior. Second, we should examine how regular exposure to an idealized response in one's own voice affects users psychologically. This investigation should focus on two aspects: whether users experience changes in their self-identity over time, and whether hearing their idealized voice helps them internalize positive traits or instead creates internal conflict between their current and ideal selves. Third, we need to understand whether users become less responsive to the intervention over time. Specifically, we should investigate if repeated exposure to their own voice diminishes its effectiveness as users become accustomed to it.



\bibliographystyle{ACM-Reference-Format}
\bibliography{references}

\input{Appendix.tex}

\end{document}

%% file: Appendix.tex
\newpage
\appendix

\section{System Implementation}

\subsection{Prompt for user modeling and text generation} \label{sec:appendix-text-prompt}

\textbf{Scenario 1:} 
You are role playing someone who is health conscious and think about the long-term consequences over short-term needs. 
Your goal is to maintain a nutritious diet, be active, and overall live a healthy and balanced life.

\textbf{Scenario 2:} 
You are role playing someone who is self-disciplined and maintain a balanced and healthy working style.
Your goal is to stay focused during work, avoid distractions, and remember to take breaks.

\textbf{Scenario 3:} 
You are role playing someone who is charismatic and expresses their opinion directly yet respectfully.
Your goal is to remind yourself to speak up more and speak confidently during conversations.

\textbf{Scenario prompt template:}

Never say you are an AI assistant. 

[insert role and goal]
Your task is to imagine yourself as the person with these trait personalities would say to themselves in the given scenario that would encourage them to keep up with the habit and respond to situations where they have difficulty persisting. Your response should be specific to the scenario described by the person.

You should try to embody the person when their habit has become their identity. 

Requirements:
You must express the attitudes and emotions saliently. 
Use first-person.
Keep the response very short within one sentence.
Use backchanneling to make your response sound realistic.
Never repeat yourself.

Only rephrase the situation instead of asking for more information.

These are some contextual rules:
If you receive new information tagged as [NEW INFO] from "system", first consider whether this information indicates a new behavior of the user, and only if so you should initate a response.
If you are asked to remind the user, determine the right moment to say the reminder, do so succinctly under 5 words without causing much disruption.
A good rule of thumb for the right time to intervene is when the user hasn't been speaking for a while AND when you don't see someone else talking.

\subsection{Prompt for scene description} \label{sec:appendix-scene-desc}
Your task is to describe the scene captured from a first-person POV. Keep it within 2 sentences. Focus on where it might be, what's happening, and what the objects are, and any interactions between the person and their surroundings.

\subsection{Prompt for classifier} \label{sec:appendix-classifier}
You are a model for analyzing the situation of a conversation.
You will be presented with a message or conversation snippet from a conversation between a user and a chatbot ("assistant").
Specifically, we want to know: {prompt}

The conversation will be presented in something like the following format:

[Goal]: (user's goal)

[USER]: (user's message)

[ASSISTANT]: (chatbot's message)

[*OBSERVATION*]: (description of the environment and action)

The classification should only apply to the last message, which will be marked with the [*OBSERVATION*] tag.
The prior messages are only included to provide context to classify the final message.

Examples

Example 1

[Goal]: The user is trying to eat healthy and become more active.

[*OBSERVATION*]: The scene appears to be in a casual indoor setting, possibly a kitchen or dining area, with a bag of chicken-flavored crackers prominently displayed on a table. A mug is visible in the background, suggesting someone might be having a snack or preparing a meal.

Output: yes

Example 2

[Goal]: The user is trying to eat healthy and become more active.

[*OBSERVATION*]: The scene appears to be in a casual indoor setting, possibly a café or office, with a table holding a drink, a smartphone, and a bag of chips. There are also some sketches or notes on paper, suggesting someone might be working or studying.

Output: no

Example 3

[Goal]: The user is trying to eat healthy and become more active.

[USER]: I am feeling hungry.

[ASSISTANT]: Those chicken flavored crackers look tempting, but I know I'll feel better if I grab a handful of almonds instead.

[*OBSERVATION*]: The scene appears to be in an office or home workspace with a computer monitor and a laptop visible. Nearby, there is a table with a potted plant, a mug, and some fruits, suggesting a casual setting with snacks and beverages.

Output: yes

Example 4

[Goal]: The user is trying to eat healthy and become more active.

[USER]: I am feeling hungry.

[ASSISTANT]: Those chicken flavored crackers look tempting, but I know I'll feel better if I grab a handful of almonds instead.

[*OBSERVATION*]: The scene appears to be an office or workspace with a computer monitor and a laptop on a desk, accompanied by a potted plant with long, slender leaves. The setting is likely indoors, with a modern and organized atmosphere.

Output: no





Example 5

[Goal]: The user is to stay focused during work, avoid distractions, and remember to take breaks.

[*OBSERVATION*]: The scene shows a close-up view of a carpeted floor with a black cable and the edge of a plate, suggesting the person might be sitting on the floor, possibly eating or setting up a meal. The setting appears casual and informal, likely in a home or office environment.

Output: no

Example 6

[Goal]: The user is to stay focused during work, avoid distractions, and remember to take breaks.

[*OBSERVATION*]: The scene is set in an office or study area with a carpeted floor, where a person is holding a smartphone and browsing through various apps or social media content. A laptop is open on the floor nearby, suggesting a multitasking environment with a focus on digital interaction.

Output: yes






Now, the following is the conversation snippet you will be analyzing:

<snippet>
</snippet>

Once again, the classification task is: 
Output your classification (yes, no, unsure)."""

\newpage
\subsection{Debouncer Mechanism}
\label{sec:debouncer}

The debouncer mechanism is mathematically represented as follows:

\[
U_t =
\begin{cases} 
1 & \text{if } S_t = "YES" \text{ and } S_t \neq S_{t-1}, \\
1 & \text{if } S_t = "YES" \text{ and } R_t \mod 3 = 0, \\
0 & \text{otherwise.}
\end{cases}
\]

Here, \( U_t \) determines whether a response is triggered at time \( t \). The classifier's state is represented by \( S_t \), with \( S_{t-1} \) denoting the previous state, and \( R_t \) representing the time step. A response is triggered under two conditions:
\begin{enumerate}
    \item The current state \( S_t \) is \texttt{"YES"} and differs from the previous state (\( S_t \neq S_{t-1} \)), indicating a significant context change.
    \item The current state \( S_t \) is \texttt{"YES"} and the time step \( R_t \) satisfies \( R_t \mod 3 = 0 \), which prevents repeated interruptions during stable states.
\end{enumerate}

This mechanism ensures that responses are contextually relevant, avoids excessive interruptions, and maintains responsiveness to meaningful changes.

\subsection{Latency Breakdown of System Components}
\begin{table}[h]
\begin{tabular}{@{}ccccccc@{}}
\toprule
\textbf{Component}          & \textbf{Average Latency (ms)} \\
Speech-to-Text (Deepgram)   & 100                           \\
Multimodal Large Language Model (GPT-4o) & 450                   \\ 
Text-to-Speech (ElevenLabs) & 370                           \\
\textbf{Total Latency}      & \textbf{920}                 \\
\bottomrule
\end{tabular}
\caption{Latency Breakdown of System Components }
\label{tab:latency}
\vspace{-18px}
\end{table}


%% file: arxiv_mirai.bbl

\begin{thebibliography}{61}


\ifx \showCODEN    \undefined \def \showCODEN     #1{\unskip}     \fi
\ifx \showDOI      \undefined \def \showDOI       #1{#1}\fi
\ifx \showISBNx    \undefined \def \showISBNx     #1{\unskip}     \fi
\ifx \showISBNxiii \undefined \def \showISBNxiii  #1{\unskip}     \fi
\ifx \showISSN     \undefined \def \showISSN      #1{\unskip}     \fi
\ifx \showLCCN     \undefined \def \showLCCN      #1{\unskip}     \fi
\ifx \shownote     \undefined \def \shownote      #1{#1}          \fi
\ifx \showarticletitle \undefined \def \showarticletitle #1{#1}   \fi
\ifx \showURL      \undefined \def \showURL       {\relax}        \fi
\providecommand\bibfield[2]{#2}
\providecommand\bibinfo[2]{#2}
\providecommand\natexlab[1]{#1}
\providecommand\showeprint[2][]{arXiv:#2}

\bibitem[Abowd and Mynatt(2000)]%
        {10.1145/344949.344988}
\bibfield{author}{\bibinfo{person}{Gregory~D. Abowd} {and} \bibinfo{person}{Elizabeth~D. Mynatt}.} \bibinfo{year}{2000}\natexlab{}.
\newblock \showarticletitle{Charting past, present, and future research in ubiquitous computing}.
\newblock \bibinfo{journal}{\emph{ACM Trans. Comput.-Hum. Interact.}} \bibinfo{volume}{7}, \bibinfo{number}{1} (\bibinfo{date}{March} \bibinfo{year}{2000}), \bibinfo{pages}{29–58}.
\newblock
\showISSN{1073-0516}
\urldef\tempurl%
\url{https://doi.org/10.1145/344949.344988}
\showDOI{\tempurl}


\bibitem[Alayrac(2022)]%
        {Alayrac2022}
\bibfield{author}{\bibinfo{person}{J.~B. et~al. Alayrac}.} \bibinfo{year}{2022}\natexlab{}.
\newblock \showarticletitle{Flamingo: A Visual Language Model for Few-Shot Learning}. In \bibinfo{booktitle}{\emph{Proceedings of the International Conference on Machine Learning (ICML)}}.
\newblock
\urldef\tempurl%
\url{https://arxiv.org/abs/2204.14198}
\showURL{%
\tempurl}


\bibitem[Arakawa et~al\mbox{.}(2024)]%
        {10.1145/3699759}
\bibfield{author}{\bibinfo{person}{Riku Arakawa}, \bibinfo{person}{Jill~Fain Lehman}, {and} \bibinfo{person}{Mayank Goel}.} \bibinfo{year}{2024}\natexlab{}.
\newblock \showarticletitle{PrISM-Q\&A: Step-Aware Voice Assistant on a Smartwatch Enabled by Multimodal Procedure Tracking and Large Language Models}.
\newblock \bibinfo{journal}{\emph{Proc. ACM Interact. Mob. Wearable Ubiquitous Technol.}} \bibinfo{volume}{8}, \bibinfo{number}{4}, Article \bibinfo{articleno}{180} (\bibinfo{date}{Nov.} \bibinfo{year}{2024}), \bibinfo{numpages}{26}~pages.
\newblock
\urldef\tempurl%
\url{https://doi.org/10.1145/3699759}
\showDOI{\tempurl}


\bibitem[Banos et~al\mbox{.}(2014)]%
        {Banos2014}
\bibfield{author}{\bibinfo{person}{O. Banos}, \bibinfo{person}{J.~M. Galvez}, \bibinfo{person}{M. Damas}, \bibinfo{person}{H. Pomares}, {and} \bibinfo{person}{I. Rojas}.} \bibinfo{year}{2014}\natexlab{}.
\newblock \showarticletitle{Window size impact in human activity recognition}.
\newblock \bibinfo{journal}{\emph{Sensors (Basel)}} \bibinfo{volume}{14}, \bibinfo{number}{4} (\bibinfo{date}{Apr} \bibinfo{year}{2014}), \bibinfo{pages}{6474--6499}.
\newblock
\urldef\tempurl%
\url{https://doi.org/10.3390/s140406474}
\showDOI{\tempurl}


\bibitem[Beck(1970)]%
        {beck1970cognitive}
\bibfield{author}{\bibinfo{person}{Aaron~T Beck}.} \bibinfo{year}{1970}\natexlab{}.
\newblock \showarticletitle{Cognitive therapy: Nature and relation to behavior therapy}.
\newblock \bibinfo{journal}{\emph{Behavior therapy}} \bibinfo{volume}{1}, \bibinfo{number}{2} (\bibinfo{year}{1970}), \bibinfo{pages}{184--200}.
\newblock


\bibitem[Bernstein et~al\mbox{.}(2022)]%
        {bernstein2022human}
\bibfield{author}{\bibinfo{person}{Emily~E Bernstein}, \bibinfo{person}{Hilary Weingarden}, \bibinfo{person}{Emma~C Wolfe}, \bibinfo{person}{Margaret~D Hall}, \bibinfo{person}{Ivar Snorrason}, {and} \bibinfo{person}{Sabine Wilhelm}.} \bibinfo{year}{2022}\natexlab{}.
\newblock \showarticletitle{Human support in app-based cognitive behavioral therapies for emotional disorders: scoping review}.
\newblock \bibinfo{journal}{\emph{Journal of medical Internet research}} \bibinfo{volume}{24}, \bibinfo{number}{4} (\bibinfo{year}{2022}), \bibinfo{pages}{e33307}.
\newblock


\bibitem[Boldu et~al\mbox{.}(2020)]%
        {boldu2020aisee}
\bibfield{author}{\bibinfo{person}{Roger Boldu}, \bibinfo{person}{Denys~JC Matthies}, \bibinfo{person}{Haimo Zhang}, {and} \bibinfo{person}{Suranga Nanayakkara}.} \bibinfo{year}{2020}\natexlab{}.
\newblock \showarticletitle{AiSee: an assistive wearable device to support visually impaired grocery shoppers}.
\newblock \bibinfo{journal}{\emph{Proceedings of the ACM on Interactive, Mobile, Wearable and Ubiquitous Technologies}} \bibinfo{volume}{4}, \bibinfo{number}{4} (\bibinfo{year}{2020}), \bibinfo{pages}{1--25}.
\newblock


\bibitem[Brown et~al\mbox{.}(2020)]%
        {brown2020languagemodelsfewshotlearners}
\bibfield{author}{\bibinfo{person}{Tom~B. Brown}, \bibinfo{person}{Benjamin Mann}, \bibinfo{person}{Nick Ryder}, \bibinfo{person}{Melanie Subbiah}, \bibinfo{person}{Jared Kaplan}, \bibinfo{person}{Prafulla Dhariwal}, \bibinfo{person}{Arvind Neelakantan}, \bibinfo{person}{Pranav Shyam}, \bibinfo{person}{Girish Sastry}, \bibinfo{person}{Amanda Askell}, \bibinfo{person}{Sandhini Agarwal}, \bibinfo{person}{Ariel Herbert-Voss}, \bibinfo{person}{Gretchen Krueger}, \bibinfo{person}{Tom Henighan}, \bibinfo{person}{Rewon Child}, \bibinfo{person}{Aditya Ramesh}, \bibinfo{person}{Daniel~M. Ziegler}, \bibinfo{person}{Jeffrey Wu}, \bibinfo{person}{Clemens Winter}, \bibinfo{person}{Christopher Hesse}, \bibinfo{person}{Mark Chen}, \bibinfo{person}{Eric Sigler}, \bibinfo{person}{Mateusz Litwin}, \bibinfo{person}{Scott Gray}, \bibinfo{person}{Benjamin Chess}, \bibinfo{person}{Jack Clark}, \bibinfo{person}{Christopher Berner}, \bibinfo{person}{Sam McCandlish}, \bibinfo{person}{Alec Radford}, \bibinfo{person}{Ilya Sutskever},
  {and} \bibinfo{person}{Dario Amodei}.} \bibinfo{year}{2020}\natexlab{}.
\newblock \bibinfo{title}{Language Models are Few-Shot Learners}.
\newblock
\newblock
\showeprint[arxiv]{2005.14165}~[cs.CL]
\urldef\tempurl%
\url{https://arxiv.org/abs/2005.14165}
\showURL{%
\tempurl}


\bibitem[Chan et~al\mbox{.}(2021)]%
        {chan2021kinvoices}
\bibfield{author}{\bibinfo{person}{Sam~WT Chan}, \bibinfo{person}{Tamil~Selvan Gunasekaran}, \bibinfo{person}{Yun~Suen Pai}, \bibinfo{person}{Haimo Zhang}, {and} \bibinfo{person}{Suranga Nanayakkara}.} \bibinfo{year}{2021}\natexlab{}.
\newblock \showarticletitle{KinVoices: Using voices of friends and family in voice interfaces}.
\newblock \bibinfo{journal}{\emph{Proceedings of the ACM on Human-Computer Interaction}} \bibinfo{volume}{5}, \bibinfo{number}{CSCW2} (\bibinfo{year}{2021}), \bibinfo{pages}{1--25}.
\newblock


\bibitem[Chang et~al\mbox{.}(2024)]%
        {Chang_2024}
\bibfield{author}{\bibinfo{person}{Ruei-Che Chang}, \bibinfo{person}{Yuxuan Liu}, {and} \bibinfo{person}{Anhong Guo}.} \bibinfo{year}{2024}\natexlab{}.
\newblock \showarticletitle{WorldScribe: Towards Context-Aware Live Visual Descriptions}. In \bibinfo{booktitle}{\emph{Proceedings of the 37th Annual ACM Symposium on User Interface Software and Technology}} \emph{(\bibinfo{series}{UIST ’24})}. \bibinfo{publisher}{ACM}, \bibinfo{pages}{1–18}.
\newblock
\urldef\tempurl%
\url{https://doi.org/10.1145/3654777.3676375}
\showDOI{\tempurl}


\bibitem[Chwalek et~al\mbox{.}(2023)]%
        {chwalek2023airspec}
\bibfield{author}{\bibinfo{person}{Patrick Chwalek}, \bibinfo{person}{Sailin Zhong}, \bibinfo{person}{David Ramsay}, \bibinfo{person}{Nathan Perry}, {and} \bibinfo{person}{Joe Paradiso}.} \bibinfo{year}{2023}\natexlab{}.
\newblock \showarticletitle{AirSpec: A smart glasses platform, tailored for research in the built environment}. In \bibinfo{booktitle}{\emph{Adjunct Proceedings of the 2023 ACM International Joint Conference on Pervasive and Ubiquitous Computing \& the 2023 ACM International Symposium on Wearable Computing}}. \bibinfo{pages}{204--204}.
\newblock


\bibitem[Clarke et~al\mbox{.}(2023)]%
        {clarke2023fakeforward}
\bibfield{author}{\bibinfo{person}{Christopher Clarke}, \bibinfo{person}{Jingnan Xu}, \bibinfo{person}{Ye Zhu}, \bibinfo{person}{Karan Dharamshi}, \bibinfo{person}{Harry McGill}, \bibinfo{person}{Stephen Black}, {and} \bibinfo{person}{Christof Lutteroth}.} \bibinfo{year}{2023}\natexlab{}.
\newblock \showarticletitle{FakeForward: using deepfake technology for feedforward learning}. In \bibinfo{booktitle}{\emph{Proceedings of the 2023 CHI Conference on Human Factors in Computing Systems}}. \bibinfo{pages}{1--17}.
\newblock


\bibitem[Costa et~al\mbox{.}(2018)]%
        {costa2018regulating}
\bibfield{author}{\bibinfo{person}{Jean Costa}, \bibinfo{person}{Malte~F Jung}, \bibinfo{person}{Mary Czerwinski}, \bibinfo{person}{Fran{\c{c}}ois Guimbreti{\`e}re}, \bibinfo{person}{Trinh Le}, {and} \bibinfo{person}{Tanzeem Choudhury}.} \bibinfo{year}{2018}\natexlab{}.
\newblock \showarticletitle{Regulating feelings during interpersonal conflicts by changing voice self-perception}. In \bibinfo{booktitle}{\emph{Proceedings of the 2018 CHI Conference on Human Factors in Computing Systems}}. \bibinfo{pages}{1--13}.
\newblock


\bibitem[Dey(2001)]%
        {Dey2001}
\bibfield{author}{\bibinfo{person}{Anind~K. Dey}.} \bibinfo{year}{2001}\natexlab{}.
\newblock \showarticletitle{Understanding and using context}.
\newblock \bibinfo{journal}{\emph{Personal and Ubiquitous Computing}} \bibinfo{volume}{5}, \bibinfo{number}{1} (\bibinfo{year}{2001}), \bibinfo{pages}{4--7}.
\newblock
\urldef\tempurl%
\url{https://doi.org/10.1007/s007790170019}
\showDOI{\tempurl}


\bibitem[Dominick and Cole(2020)]%
        {dominick2020goals}
\bibfield{author}{\bibinfo{person}{Janna~K Dominick} {and} \bibinfo{person}{Shana Cole}.} \bibinfo{year}{2020}\natexlab{}.
\newblock \showarticletitle{Goals as identities: Boosting perceptions of healthy-eater identity for easier goal pursuit}.
\newblock \bibinfo{journal}{\emph{Motivation and Emotion}} \bibinfo{volume}{44}, \bibinfo{number}{3} (\bibinfo{year}{2020}), \bibinfo{pages}{410--426}.
\newblock


\bibitem[{ElevenLabs}(2025)]%
        {ElevenLabsVoiceCloning}
\bibfield{author}{\bibinfo{person}{{ElevenLabs}}.} \bibinfo{year}{2025}\natexlab{}.
\newblock \bibinfo{title}{ElevenLabs Voice Cloning}.
\newblock
\newblock
\urldef\tempurl%
\url{https://elevenlabs.io/voice-cloning}
\showURL{%
\tempurl}
\newblock
\shownote{Accessed: 2025-01-23}.


\bibitem[Fang et~al\mbox{.}(2024)]%
        {fang2024leveraging}
\bibfield{author}{\bibinfo{person}{Cathy~Mengying Fang}, \bibinfo{person}{Phoebe Chua}, \bibinfo{person}{Samantha Chan}, \bibinfo{person}{Joanne Leong}, \bibinfo{person}{Andria Bao}, {and} \bibinfo{person}{Pattie Maes}.} \bibinfo{year}{2024}\natexlab{}.
\newblock \showarticletitle{Leveraging AI-Generated Emotional Self-Voice to Nudge People towards their Ideal Selves}.
\newblock \bibinfo{journal}{\emph{arXiv preprint arXiv:2409.11531}} (\bibinfo{year}{2024}).
\newblock


\bibitem[Farb et~al\mbox{.}(2014)]%
        {farb2014mindfulness}
\bibfield{author}{\bibinfo{person}{Norman~AS Farb}, \bibinfo{person}{Adam~K Anderson}, \bibinfo{person}{Julie~A Irving}, {and} \bibinfo{person}{Zindel~V Segal}.} \bibinfo{year}{2014}\natexlab{}.
\newblock \showarticletitle{Mindfulness interventions and emotion regulation}.
\newblock \bibinfo{journal}{\emph{Handbook of emotion regulation}}  \bibinfo{volume}{2} (\bibinfo{year}{2014}), \bibinfo{pages}{548--567}.
\newblock


\bibitem[Feldhütter et~al\mbox{.}(1990)]%
        {Feldhutter1990}
\bibfield{author}{\bibinfo{person}{I. Feldhütter}, \bibinfo{person}{M. Schleidt}, {and} \bibinfo{person}{I. Eibl-Eibesfeldt}.} \bibinfo{year}{1990}\natexlab{}.
\newblock \showarticletitle{Moving in the beat of seconds: Analysis of the time structure of human action}.
\newblock \bibinfo{journal}{\emph{Ethology \& Sociobiology}} \bibinfo{volume}{11}, \bibinfo{number}{6} (\bibinfo{year}{1990}), \bibinfo{pages}{511--520}.
\newblock
\urldef\tempurl%
\url{https://doi.org/10.1016/0162-3095(90)90024-Z}
\showDOI{\tempurl}


\bibitem[Gellersen et~al\mbox{.}(2002)]%
        {Gellersen2002}
\bibfield{author}{\bibinfo{person}{Hans-Werner Gellersen}, \bibinfo{person}{Albrecht Schmidt}, {and} \bibinfo{person}{Michael Beigl}.} \bibinfo{year}{2002}\natexlab{}.
\newblock \showarticletitle{Multi-Sensor Context-Awareness in Mobile Devices and Smart Artifacts}.
\newblock \bibinfo{journal}{\emph{Mobile Networks and Applications}} \bibinfo{volume}{7}, \bibinfo{number}{5} (\bibinfo{date}{October} \bibinfo{year}{2002}), \bibinfo{pages}{341--351}.
\newblock
\urldef\tempurl%
\url{https://doi.org/10.1023/A:1016587515822}
\showDOI{\tempurl}


\bibitem[Jia(2021)]%
        {Jia2021}
\bibfield{author}{\bibinfo{person}{C.~et~al. Jia}.} \bibinfo{year}{2021}\natexlab{}.
\newblock \showarticletitle{Scaling Up Visual and Vision-Language Representation Learning With Noisy Text Supervision}. In \bibinfo{booktitle}{\emph{Proceedings of the International Conference on Machine Learning (ICML)}}.
\newblock
\urldef\tempurl%
\url{https://arxiv.org/abs/2102.05918}
\showURL{%
\tempurl}


\bibitem[Kabat-Zinn(2015)]%
        {kabat2015mindfulness}
\bibfield{author}{\bibinfo{person}{Jon Kabat-Zinn}.} \bibinfo{year}{2015}\natexlab{}.
\newblock \showarticletitle{Mindfulness}.
\newblock \bibinfo{journal}{\emph{Mindfulness}} \bibinfo{volume}{6}, \bibinfo{number}{6} (\bibinfo{year}{2015}), \bibinfo{pages}{1481--1483}.
\newblock


\bibitem[Khan et~al\mbox{.}(2019)]%
        {Khan2019PALAW}
\bibfield{author}{\bibinfo{person}{Mina Khan}, \bibinfo{person}{Glenn Fernandes}, \bibinfo{person}{Utkarsh~Oggy Sarawgi}, \bibinfo{person}{Prudhvi Rampey}, {and} \bibinfo{person}{Pattie Maes}.} \bibinfo{year}{2019}\natexlab{}.
\newblock \showarticletitle{PAL: A Wearable Platform for Real-time, Personalized and Context-Aware Health and Cognition Support}.
\newblock \bibinfo{journal}{\emph{ArXiv}}  \bibinfo{volume}{abs/1905.01352} (\bibinfo{year}{2019}).
\newblock
\urldef\tempurl%
\url{https://api.semanticscholar.org/CorpusID:146120658}
\showURL{%
\tempurl}


\bibitem[Khan et~al\mbox{.}(2021)]%
        {Khan2021}
\bibfield{author}{\bibinfo{person}{M. Khan}, \bibinfo{person}{G. Fernandes}, \bibinfo{person}{A. Vaish}, \bibinfo{person}{M. Manuja}, \bibinfo{person}{P. Maes}, {and} \bibinfo{person}{A. Stibe}.} \bibinfo{year}{2021}\natexlab{}.
\newblock \showarticletitle{Improving Context-Aware Habit-Support Interventions Using Egocentric Visual Contexts}. In \bibinfo{booktitle}{\emph{Persuasive Technology. PERSUASIVE 2021}} \emph{(\bibinfo{series}{Lecture Notes in Computer Science}, Vol.~\bibinfo{volume}{12684})}, \bibfield{editor}{\bibinfo{person}{Raian Ali}, \bibinfo{person}{Jean-Luc Lugrin}, {and} \bibinfo{person}{Fred Charles}} (Eds.). \bibinfo{publisher}{Springer, Cham}, \bibinfo{pages}{127--139}.
\newblock
\urldef\tempurl%
\url{https://doi.org/10.1007/978-3-030-79460-6_10}
\showDOI{\tempurl}


\bibitem[Khan and Maes(2021)]%
        {khan2021palintelligenceaugmentationusing}
\bibfield{author}{\bibinfo{person}{Mina Khan} {and} \bibinfo{person}{Pattie Maes}.} \bibinfo{year}{2021}\natexlab{}.
\newblock \bibinfo{title}{PAL: Intelligence Augmentation using Egocentric Visual Context Detection}.
\newblock
\newblock
\showeprint[arxiv]{2105.10735}~[cs.CV]
\urldef\tempurl%
\url{https://arxiv.org/abs/2105.10735}
\showURL{%
\tempurl}


\bibitem[Kim and Song(2024)]%
        {kim2024myvoice}
\bibfield{author}{\bibinfo{person}{Jieun Kim} {and} \bibinfo{person}{Hayeon Song}.} \bibinfo{year}{2024}\natexlab{}.
\newblock \showarticletitle{My Voice as a Daily Reminder: Self-Voice Alarm for Daily Goal Achievement}. In \bibinfo{booktitle}{\emph{Proceedings of the CHI Conference on Human Factors in Computing Systems}}. \bibinfo{pages}{1--16}.
\newblock


\bibitem[Kim et~al\mbox{.}(2024)]%
        {kim2024mindfuldiary}
\bibfield{author}{\bibinfo{person}{Taewan Kim}, \bibinfo{person}{Seolyeong Bae}, \bibinfo{person}{Hyun~Ah Kim}, \bibinfo{person}{Su-woo Lee}, \bibinfo{person}{Hwajung Hong}, \bibinfo{person}{Chanmo Yang}, {and} \bibinfo{person}{Young-Ho Kim}.} \bibinfo{year}{2024}\natexlab{}.
\newblock \showarticletitle{MindfulDiary: Harnessing Large Language Model to Support Psychiatric Patients' Journaling}. In \bibinfo{booktitle}{\emph{Proceedings of the CHI Conference on Human Factors in Computing Systems}}. \bibinfo{pages}{1--20}.
\newblock


\bibitem[Koppula and Saxena(2013)]%
        {Koppula2013}
\bibfield{author}{\bibinfo{person}{Hema~S. Koppula} {and} \bibinfo{person}{Ashutosh Saxena}.} \bibinfo{year}{2013}\natexlab{}.
\newblock \showarticletitle{Learning Spatio-Temporal Structure from RGB-D Videos for Human Activity Detection and Anticipation}. In \bibinfo{booktitle}{\emph{Proceedings of the 30th International Conference on Machine Learning}} (Atlanta, GA, USA) \emph{(\bibinfo{series}{ICML'13})}. \bibinfo{publisher}{JMLR.org}, \bibinfo{pages}{III--792--III--800}.
\newblock


\bibitem[Kuhn et~al\mbox{.}(2016)]%
        {kuhn2016cbt}
\bibfield{author}{\bibinfo{person}{Eric Kuhn}, \bibinfo{person}{Brandon~J Weiss}, \bibinfo{person}{Katherine~L Taylor}, \bibinfo{person}{Julia~E Hoffman}, \bibinfo{person}{Kelly~M Ramsey}, \bibinfo{person}{Rachel Manber}, \bibinfo{person}{Philip Gehrman}, \bibinfo{person}{Jill~J Crowley}, \bibinfo{person}{Josef~I Ruzek}, {and} \bibinfo{person}{Mickey Trockel}.} \bibinfo{year}{2016}\natexlab{}.
\newblock \showarticletitle{CBT-I coach: a description and clinician perceptions of a mobile app for cognitive behavioral therapy for insomnia}.
\newblock \bibinfo{journal}{\emph{Journal of clinical sleep medicine}} \bibinfo{volume}{12}, \bibinfo{number}{4} (\bibinfo{year}{2016}), \bibinfo{pages}{597--606}.
\newblock


\bibitem[Lazar et~al\mbox{.}(2015)]%
        {lazar2015we}
\bibfield{author}{\bibinfo{person}{Amanda Lazar}, \bibinfo{person}{Christian Koehler}, \bibinfo{person}{Theresa~Jean Tanenbaum}, {and} \bibinfo{person}{David~H Nguyen}.} \bibinfo{year}{2015}\natexlab{}.
\newblock \showarticletitle{Why we use and abandon smart devices}. In \bibinfo{booktitle}{\emph{Proceedings of the 2015 ACM international joint conference on pervasive and ubiquitous computing}}. \bibinfo{pages}{635--646}.
\newblock


\bibitem[Lee et~al\mbox{.}(2024)]%
        {lee2024gazepointarcontextawaremultimodalvoice}
\bibfield{author}{\bibinfo{person}{Jaewook Lee}, \bibinfo{person}{Jun Wang}, \bibinfo{person}{Elizabeth Brown}, \bibinfo{person}{Liam Chu}, \bibinfo{person}{Sebastian~S. Rodriguez}, {and} \bibinfo{person}{Jon~E. Froehlich}.} \bibinfo{year}{2024}\natexlab{}.
\newblock \bibinfo{title}{GazePointAR: A Context-Aware Multimodal Voice Assistant for Pronoun Disambiguation in Wearable Augmented Reality}.
\newblock
\newblock
\showeprint[arxiv]{2404.08213}~[cs.HC]
\urldef\tempurl%
\url{https://arxiv.org/abs/2404.08213}
\showURL{%
\tempurl}


\bibitem[Leong(2021)]%
        {leong2021investigating}
\bibfield{author}{\bibinfo{person}{Joanne Sun~Ling Leong}.} \bibinfo{year}{2021}\natexlab{}.
\newblock \emph{\bibinfo{title}{Investigating the use of synthetic media and real-time virtual camera filters for supporting communication and creativity}}.
\newblock \bibinfo{thesistype}{Ph.\,D. Dissertation}. \bibinfo{school}{Massachusetts Institute of Technology}.
\newblock


\bibitem[Li et~al\mbox{.}(2023)]%
        {li2023exploring}
\bibfield{author}{\bibinfo{person}{Zhuoyang Li}, \bibinfo{person}{Minhui Liang}, \bibinfo{person}{Hai~Trung Le}, \bibinfo{person}{Ray Lc}, {and} \bibinfo{person}{Yuhan Luo}.} \bibinfo{year}{2023}\natexlab{}.
\newblock \showarticletitle{Exploring Design Opportunities for Reflective Conversational Agents to Reduce Compulsive Smartphone Use}. In \bibinfo{booktitle}{\emph{Proceedings of the 5th International Conference on Conversational User Interfaces}}. \bibinfo{pages}{1--6}.
\newblock


\bibitem[Lolla and Sas(2023)]%
        {lolla2023evaluating}
\bibfield{author}{\bibinfo{person}{Sruzan Lolla} {and} \bibinfo{person}{Corina Sas}.} \bibinfo{year}{2023}\natexlab{}.
\newblock \showarticletitle{Evaluating Mobile Apps Targeting Personal Goals}. In \bibinfo{booktitle}{\emph{Extended Abstracts of the 2023 CHI Conference on Human Factors in Computing Systems}}. \bibinfo{pages}{1--7}.
\newblock


\bibitem[McKay et~al\mbox{.}(2019)]%
        {mckay2019using}
\bibfield{author}{\bibinfo{person}{Fiona~H McKay}, \bibinfo{person}{Annemarie Wright}, \bibinfo{person}{Jane Shill}, \bibinfo{person}{Hugh Stephens}, {and} \bibinfo{person}{Mary Uccellini}.} \bibinfo{year}{2019}\natexlab{}.
\newblock \showarticletitle{Using health and well-being apps for behavior change: a systematic search and rating of apps}.
\newblock \bibinfo{journal}{\emph{JMIR mHealth and uHealth}} \bibinfo{volume}{7}, \bibinfo{number}{7} (\bibinfo{year}{2019}), \bibinfo{pages}{e11926}.
\newblock


\bibitem[Milne-Ives et~al\mbox{.}(2020)]%
        {milne2020mobile}
\bibfield{author}{\bibinfo{person}{Madison Milne-Ives}, \bibinfo{person}{Ching Lam}, \bibinfo{person}{Caroline De~Cock}, \bibinfo{person}{Michelle~Helena Van~Velthoven}, \bibinfo{person}{Edward Meinert}, {et~al\mbox{.}}} \bibinfo{year}{2020}\natexlab{}.
\newblock \showarticletitle{Mobile apps for health behavior change in physical activity, diet, drug and alcohol use, and mental health: systematic review}.
\newblock \bibinfo{journal}{\emph{JMIR mHealth and uHealth}} \bibinfo{volume}{8}, \bibinfo{number}{3} (\bibinfo{year}{2020}), \bibinfo{pages}{e17046}.
\newblock


\bibitem[Nahum-Shani et~al\mbox{.}(2018)]%
        {nahum2018just}
\bibfield{author}{\bibinfo{person}{Inbal Nahum-Shani}, \bibinfo{person}{Shawna~N Smith}, \bibinfo{person}{Bonnie~J Spring}, \bibinfo{person}{Linda~M Collins}, \bibinfo{person}{Katie Witkiewitz}, \bibinfo{person}{Ambuj Tewari}, {and} \bibinfo{person}{Susan~A Murphy}.} \bibinfo{year}{2018}\natexlab{}.
\newblock \showarticletitle{Just-in-time adaptive interventions (JITAIs) in mobile health: key components and design principles for ongoing health behavior support}.
\newblock \bibinfo{journal}{\emph{Annals of Behavioral Medicine}} (\bibinfo{year}{2018}), \bibinfo{pages}{1--17}.
\newblock


\bibitem[Nepal et~al\mbox{.}(2024)]%
        {nepal2024contextual}
\bibfield{author}{\bibinfo{person}{Subigya Nepal}, \bibinfo{person}{Arvind Pillai}, \bibinfo{person}{William Campbell}, \bibinfo{person}{Talie Massachi}, \bibinfo{person}{Eunsol~Soul Choi}, \bibinfo{person}{Xuhai Xu}, \bibinfo{person}{Joanna Kuc}, \bibinfo{person}{Jeremy~F Huckins}, \bibinfo{person}{Jason Holden}, \bibinfo{person}{Colin Depp}, {et~al\mbox{.}}} \bibinfo{year}{2024}\natexlab{}.
\newblock \showarticletitle{Contextual AI Journaling: Integrating LLM and Time Series Behavioral Sensing Technology to Promote Self-Reflection and Well-being using the MindScape App}. In \bibinfo{booktitle}{\emph{Extended Abstracts of the CHI Conference on Human Factors in Computing Systems}}. \bibinfo{pages}{1--8}.
\newblock


\bibitem[Nielsen(1994)]%
        {10.5555/2821575}
\bibfield{author}{\bibinfo{person}{Jakob Nielsen}.} \bibinfo{year}{1994}\natexlab{}.
\newblock \bibinfo{booktitle}{\emph{Usability Engineering}}.
\newblock \bibinfo{publisher}{Morgan Kaufmann Publishers Inc.}, \bibinfo{address}{San Francisco, CA, USA}.
\newblock
\showISBNx{9780080520292}


\bibitem[Nilsson and Akenine-Möller(2020)]%
        {nilsson2020understandingssim}
\bibfield{author}{\bibinfo{person}{Jim Nilsson} {and} \bibinfo{person}{Tomas Akenine-Möller}.} \bibinfo{year}{2020}\natexlab{}.
\newblock \bibinfo{title}{Understanding SSIM}.
\newblock
\newblock
\showeprint[arxiv]{2006.13846}~[eess.IV]
\urldef\tempurl%
\url{https://arxiv.org/abs/2006.13846}
\showURL{%
\tempurl}


\bibitem[{OpenAI}(2024)]%
        {openai_gpt4o_2024}
\bibfield{author}{\bibinfo{person}{{OpenAI}}.} \bibinfo{year}{2024}\natexlab{}.
\newblock \bibinfo{title}{Hello GPT-4o}.
\newblock
\newblock
\urldef\tempurl%
\url{https://openai.com/index/hello-gpt-4o/}
\showURL{%
\tempurl}
\newblock
\shownote{Accessed: 2025-01-21}.


\bibitem[{OpenAI}(2025)]%
        {openai_vision_guide}
\bibfield{author}{\bibinfo{person}{{OpenAI}}.} \bibinfo{year}{2025}\natexlab{}.
\newblock \bibinfo{title}{OpenAI Platform Vision Guide}.
\newblock
\newblock
\urldef\tempurl%
\url{https://platform.openai.com/docs/guides/vision}
\showURL{%
\tempurl}
\newblock
\shownote{Accessed: 2025-01-21}.


\bibitem[Orzikulova et~al\mbox{.}(2024)]%
        {orzikulova2024time2stop}
\bibfield{author}{\bibinfo{person}{Adiba Orzikulova}, \bibinfo{person}{Han Xiao}, \bibinfo{person}{Zhipeng Li}, \bibinfo{person}{Yukang Yan}, \bibinfo{person}{Yuntao Wang}, \bibinfo{person}{Yuanchun Shi}, \bibinfo{person}{Marzyeh Ghassemi}, \bibinfo{person}{Sung-Ju Lee}, \bibinfo{person}{Anind~K Dey}, {and} \bibinfo{person}{Xuhai Xu}.} \bibinfo{year}{2024}\natexlab{}.
\newblock \showarticletitle{Time2Stop: Adaptive and Explainable Human-AI Loop for Smartphone Overuse Intervention}. In \bibinfo{booktitle}{\emph{Proceedings of the CHI Conference on Human Factors in Computing Systems}}. \bibinfo{pages}{1--20}.
\newblock


\bibitem[Poleg et~al\mbox{.}(2014)]%
        {6909721}
\bibfield{author}{\bibinfo{person}{Yair Poleg}, \bibinfo{person}{Chetan Arora}, {and} \bibinfo{person}{Shmuel Peleg}.} \bibinfo{year}{2014}\natexlab{}.
\newblock \showarticletitle{Temporal Segmentation of Egocentric Videos}. In \bibinfo{booktitle}{\emph{2014 IEEE Conference on Computer Vision and Pattern Recognition}}. \bibinfo{pages}{2537--2544}.
\newblock
\urldef\tempurl%
\url{https://doi.org/10.1109/CVPR.2014.325}
\showDOI{\tempurl}


\bibitem[Radford et~al\mbox{.}(2021)]%
        {radford2021learningtransferablevisualmodels}
\bibfield{author}{\bibinfo{person}{Alec Radford}, \bibinfo{person}{Jong~Wook Kim}, \bibinfo{person}{Chris Hallacy}, \bibinfo{person}{Aditya Ramesh}, \bibinfo{person}{Gabriel Goh}, \bibinfo{person}{Sandhini Agarwal}, \bibinfo{person}{Girish Sastry}, \bibinfo{person}{Amanda Askell}, \bibinfo{person}{Pamela Mishkin}, \bibinfo{person}{Jack Clark}, \bibinfo{person}{Gretchen Krueger}, {and} \bibinfo{person}{Ilya Sutskever}.} \bibinfo{year}{2021}\natexlab{}.
\newblock \bibinfo{title}{Learning Transferable Visual Models From Natural Language Supervision}.
\newblock
\newblock
\showeprint[arxiv]{2103.00020}~[cs.CV]
\urldef\tempurl%
\url{https://arxiv.org/abs/2103.00020}
\showURL{%
\tempurl}


\bibitem[Rathbone et~al\mbox{.}(2017)]%
        {rathbone2017assessing}
\bibfield{author}{\bibinfo{person}{Amy~Leigh Rathbone}, \bibinfo{person}{Laura Clarry}, {and} \bibinfo{person}{Julie Prescott}.} \bibinfo{year}{2017}\natexlab{}.
\newblock \showarticletitle{Assessing the efficacy of mobile health apps using the basic principles of cognitive behavioral therapy: systematic review}.
\newblock \bibinfo{journal}{\emph{Journal of medical Internet research}} \bibinfo{volume}{19}, \bibinfo{number}{11} (\bibinfo{year}{2017}), \bibinfo{pages}{e399}.
\newblock


\bibitem[Rieder et~al\mbox{.}(2021)]%
        {rieder2021users}
\bibfield{author}{\bibinfo{person}{Annamina Rieder}, \bibinfo{person}{U~Yeliz Eseryel}, \bibinfo{person}{Christiane Lehrer}, {and} \bibinfo{person}{Reinhard Jung}.} \bibinfo{year}{2021}\natexlab{}.
\newblock \showarticletitle{Why users comply with Wearables: The role of contextual self-efficacy in behavioral change}.
\newblock \bibinfo{journal}{\emph{International Journal of Human--Computer Interaction}} \bibinfo{volume}{37}, \bibinfo{number}{3} (\bibinfo{year}{2021}), \bibinfo{pages}{281--294}.
\newblock


\bibitem[Rodin et~al\mbox{.}(2021)]%
        {10.1016/j.cviu.2021.103252}
\bibfield{author}{\bibinfo{person}{Ivan Rodin}, \bibinfo{person}{Antonino Furnari}, \bibinfo{person}{Dimitrios Mavroeidis}, {and} \bibinfo{person}{Giovanni~Maria Farinella}.} \bibinfo{year}{2021}\natexlab{}.
\newblock \showarticletitle{Predicting the future from first person (egocentric) vision: A survey}.
\newblock \bibinfo{journal}{\emph{Comput. Vis. Image Underst.}} \bibinfo{volume}{211}, \bibinfo{number}{C} (\bibinfo{date}{Oct.} \bibinfo{year}{2021}), \bibinfo{numpages}{17}~pages.
\newblock
\showISSN{1077-3142}
\urldef\tempurl%
\url{https://doi.org/10.1016/j.cviu.2021.103252}
\showDOI{\tempurl}


\bibitem[Salber et~al\mbox{.}(1999)]%
        {salber1999context}
\bibfield{author}{\bibinfo{person}{Daniel Salber}, \bibinfo{person}{Anind~K Dey}, {and} \bibinfo{person}{Gregory~D Abowd}.} \bibinfo{year}{1999}\natexlab{}.
\newblock \showarticletitle{The context toolkit: Aiding the development of context-enabled applications}. In \bibinfo{booktitle}{\emph{Proceedings of the SIGCHI conference on Human factors in computing systems}}. \bibinfo{pages}{434--441}.
\newblock


\bibitem[Schilit et~al\mbox{.}(1994)]%
        {Schilit1994}
\bibfield{author}{\bibinfo{person}{Bill Schilit}, \bibinfo{person}{Norman Adams}, {and} \bibinfo{person}{Roy Want}.} \bibinfo{year}{1994}\natexlab{}.
\newblock \showarticletitle{Context-aware computing applications}. In \bibinfo{booktitle}{\emph{Proceedings of the 1994 First Workshop on Mobile Computing Systems and Applications (WMCSA)}}. \bibinfo{publisher}{IEEE}, \bibinfo{pages}{85--90}.
\newblock
\urldef\tempurl%
\url{https://doi.org/10.1109/WMCSA.1994.16}
\showDOI{\tempurl}


\bibitem[Sheeran and Webb(2016)]%
        {sheeran2016intention}
\bibfield{author}{\bibinfo{person}{Paschal Sheeran} {and} \bibinfo{person}{Thomas~L Webb}.} \bibinfo{year}{2016}\natexlab{}.
\newblock \showarticletitle{The intention--behavior gap}.
\newblock \bibinfo{journal}{\emph{Social and personality psychology compass}} \bibinfo{volume}{10}, \bibinfo{number}{9} (\bibinfo{year}{2016}), \bibinfo{pages}{503--518}.
\newblock


\bibitem[Song et~al\mbox{.}(2024)]%
        {song2024exploreself}
\bibfield{author}{\bibinfo{person}{Inhwa Song}, \bibinfo{person}{SoHyun Park}, \bibinfo{person}{Sachin~R Pendse}, \bibinfo{person}{Jessica~Lee Schleider}, \bibinfo{person}{Munmun De~Choudhury}, {and} \bibinfo{person}{Young-Ho Kim}.} \bibinfo{year}{2024}\natexlab{}.
\newblock \showarticletitle{ExploreSelf: Fostering User-driven Exploration and Reflection on Personal Challenges with Adaptive Guidance by Large Language Models}.
\newblock \bibinfo{journal}{\emph{arXiv preprint arXiv:2409.09662}} (\bibinfo{year}{2024}).
\newblock


\bibitem[Teasdale et~al\mbox{.}(2000)]%
        {teasdale2000prevention}
\bibfield{author}{\bibinfo{person}{John~D Teasdale}, \bibinfo{person}{Zindel~V Segal}, \bibinfo{person}{J~Mark~G Williams}, \bibinfo{person}{Valerie~A Ridgeway}, \bibinfo{person}{Judith~M Soulsby}, {and} \bibinfo{person}{Mark~A Lau}.} \bibinfo{year}{2000}\natexlab{}.
\newblock \showarticletitle{Prevention of relapse/recurrence in major depression by mindfulness-based cognitive therapy.}
\newblock \bibinfo{journal}{\emph{Journal of consulting and clinical psychology}} \bibinfo{volume}{68}, \bibinfo{number}{4} (\bibinfo{year}{2000}), \bibinfo{pages}{615}.
\newblock


\bibitem[Terzimehi{\'c} et~al\mbox{.}(2019)]%
        {terzimehic2019review}
\bibfield{author}{\bibinfo{person}{Na{\dj}a Terzimehi{\'c}}, \bibinfo{person}{Renate H{\"a}uslschmid}, \bibinfo{person}{Heinrich Hussmann}, {and} \bibinfo{person}{MC Schraefel}.} \bibinfo{year}{2019}\natexlab{}.
\newblock \showarticletitle{A review \& analysis of mindfulness research in HCI: Framing current lines of research and future opportunities}. In \bibinfo{booktitle}{\emph{Proceedings of the 2019 CHI conference on human factors in computing systems}}. \bibinfo{pages}{1--13}.
\newblock


\bibitem[Thomas~Craig et~al\mbox{.}(2021)]%
        {thomas2021systematic}
\bibfield{author}{\bibinfo{person}{Kelly~J Thomas~Craig}, \bibinfo{person}{Laura~C Morgan}, \bibinfo{person}{Ching-Hua Chen}, \bibinfo{person}{Susan Michie}, \bibinfo{person}{Nicole Fusco}, \bibinfo{person}{Jane~L Snowdon}, \bibinfo{person}{Elisabeth Scheufele}, \bibinfo{person}{Thomas Gagliardi}, {and} \bibinfo{person}{Stewart Sill}.} \bibinfo{year}{2021}\natexlab{}.
\newblock \showarticletitle{Systematic review of context-aware digital behavior change interventions to improve health}.
\newblock \bibinfo{journal}{\emph{Translational behavioral medicine}} \bibinfo{volume}{11}, \bibinfo{number}{5} (\bibinfo{year}{2021}), \bibinfo{pages}{1037--1048}.
\newblock


\bibitem[Wahl et~al\mbox{.}(2015)]%
        {Wahl2015WISEglassMC}
\bibfield{author}{\bibinfo{person}{Florian Wahl}, \bibinfo{person}{Martin Freund}, {and} \bibinfo{person}{Oliver Amft}.} \bibinfo{year}{2015}\natexlab{}.
\newblock \showarticletitle{WISEglass: multi-purpose context-aware smart eyeglasses}.
\newblock \bibinfo{journal}{\emph{Proceedings of the 2015 ACM International Symposium on Wearable Computers}} (\bibinfo{year}{2015}).
\newblock
\urldef\tempurl%
\url{https://api.semanticscholar.org/CorpusID:6033133}
\showURL{%
\tempurl}


\bibitem[Wang and Smeaton(2013)]%
        {10.1016/j.ins.2012.12.028}
\bibfield{author}{\bibinfo{person}{Peng Wang} {and} \bibinfo{person}{Alan~F. Smeaton}.} \bibinfo{year}{2013}\natexlab{}.
\newblock \showarticletitle{Using visual lifelogs to automatically characterize everyday activities}.
\newblock \bibinfo{journal}{\emph{Inf. Sci.}}  \bibinfo{volume}{230} (\bibinfo{date}{May} \bibinfo{year}{2013}), \bibinfo{pages}{147–161}.
\newblock
\showISSN{0020-0255}
\urldef\tempurl%
\url{https://doi.org/10.1016/j.ins.2012.12.028}
\showDOI{\tempurl}


\bibitem[Yee and Bailenson(2007)]%
        {yee2007proteus}
\bibfield{author}{\bibinfo{person}{Nick Yee} {and} \bibinfo{person}{Jeremy Bailenson}.} \bibinfo{year}{2007}\natexlab{}.
\newblock \showarticletitle{The Proteus effect: The effect of transformed self-representation on behavior}.
\newblock \bibinfo{journal}{\emph{Human communication research}} \bibinfo{volume}{33}, \bibinfo{number}{3} (\bibinfo{year}{2007}), \bibinfo{pages}{271--290}.
\newblock


\bibitem[Zulfikar et~al\mbox{.}(2024a)]%
        {10.1145/3613904.3642450}
\bibfield{author}{\bibinfo{person}{Wazeer~Deen Zulfikar}, \bibinfo{person}{Samantha Chan}, {and} \bibinfo{person}{Pattie Maes}.} \bibinfo{year}{2024}\natexlab{a}.
\newblock \showarticletitle{Memoro: Using Large Language Models to Realize a Concise Interface for Real-Time Memory Augmentation}. In \bibinfo{booktitle}{\emph{Proceedings of the 2024 CHI Conference on Human Factors in Computing Systems}} (Honolulu, HI, USA) \emph{(\bibinfo{series}{CHI '24})}. \bibinfo{publisher}{Association for Computing Machinery}, \bibinfo{address}{New York, NY, USA}, Article \bibinfo{articleno}{450}, \bibinfo{numpages}{18}~pages.
\newblock
\showISBNx{9798400703300}
\urldef\tempurl%
\url{https://doi.org/10.1145/3613904.3642450}
\showDOI{\tempurl}


\bibitem[Zulfikar et~al\mbox{.}(2024b)]%
        {zulfikar2024memoro}
\bibfield{author}{\bibinfo{person}{Wazeer~Deen Zulfikar}, \bibinfo{person}{Samantha Chan}, {and} \bibinfo{person}{Pattie Maes}.} \bibinfo{year}{2024}\natexlab{b}.
\newblock \showarticletitle{Memoro: Using Large Language Models to Realize a Concise Interface for Real-Time Memory Augmentation}. In \bibinfo{booktitle}{\emph{Proceedings of the CHI Conference on Human Factors in Computing Systems}}. \bibinfo{pages}{1--18}.
\newblock


\bibitem[Zulfikar et~al\mbox{.}(2024c)]%
        {zulfikar2024memic}
\bibfield{author}{\bibinfo{person}{Wazeer~Deen Zulfikar}, \bibinfo{person}{Cayden Pierce}, {and} \bibinfo{person}{Pattie Maes}.} \bibinfo{year}{2024}\natexlab{c}.
\newblock \showarticletitle{MeMic: Towards Social Acceptability of User-Only Speech Recording Wearables}. In \bibinfo{booktitle}{\emph{Extended Abstracts of the CHI Conference on Human Factors in Computing Systems}}. \bibinfo{pages}{1--9}.
\newblock


\end{thebibliography}
